\documentclass[prl,showpacs,amsmath,amssymb,twocolumn, 10pt]{revtex4}
\usepackage{enumerate}
\usepackage{color}
\usepackage{amsthm}
\usepackage{dcolumn}
\usepackage{bm}
\usepackage{graphicx}

\begin{document}

\newtheorem{corollary}{Corollary}
\newtheorem{definition}{Definition}
\newtheorem{example}{Example}
\newtheorem{lemma}{Lemma}
\newtheorem{proposition}{Proposition}
\newtheorem{theorem}{Theorem}
\newtheorem*{fact}{Fact}
\newtheorem{property}{Property}
\newcommand{\bra}[1]{\langle #1|}
\newcommand{\ket}[1]{|#1\rangle}
\newcommand{\braket}[3]{\langle #1|#2|#3\rangle}
%%inner product
\newcommand{\ip}[2]{\langle #1|#2\rangle}
%%outer product
\newcommand{\op}[2]{|#1\rangle \langle #2|}

\newcommand{\tr}{{\rm tr}}
\newcommand{\supp}{{\it supp}}

\newcommand{\slocc}{\stackrel{\textrm{\small SLOCC}}{\longrightarrow}}
\newcommand{\comments}[1]{}
\newcommand{\W}{|W\rangle}
\newcommand{\WW}{\W^{\otimes2}}
\newcommand{\rk}{\textrm{rk}}
\newcommand{\sch}{\textrm{Sch}}
\newcommand{\EPR}{|\textrm{EPR}\rangle}
\newcommand{\TEPR}{|\Phi^3\rangle}
\newcommand{\GHZ}{|\textrm{GHZ}\rangle}
\newcommand{\defeq}{\stackrel{\textrm{def}}{=}}
\newcommand{\Span}{\mathrm{span}}
\newcommand {\E } {{\mathcal{E}}}
\newcommand{\In}{\mathrm{in}}
\newcommand{\Out}{\mathrm{out}}
\newcommand{\local}{\mathrm{local}}
\newcommand {\F } {{\mathcal{F}}}
\newcommand {\diag } {{\rm diag}}
\renewcommand{\b}{\mathcal{B}}
\newcommand{\h}{\mathcal{H}}
\renewcommand{\Re}{\mathrm{Re}}
\renewcommand{\Im}{\mathrm{Im}}
\newcommand{\Z}{\sigma_z}
\newcommand{\Clocal}{C^{(0)}_{\local}}
\title{Tripartite entanglement transformations and tensor rank}
\author{Eric Chitambar$^1$}
\email{echitamb@umich.edu}
\author{Runyao Duan$^{2}$}
\email{dry@tsinghua.edu.cn}
\author{Yaoyun Shi$^{3}$}
\email{shiyy@eecs.umich.edu} \affiliation{
$^1$Physics Department, University of Michigan, 450 Church Street, 
Ann Arbor, Michigan 48109-1040, USA.\\
$^2$State Key Laboratory
of Intelligent Technology and Systems,Tsinghua National Laboratory
for Information Science and Technology, Department of Computer
Science and Technology, Tsinghua University, Beijing 100084, China
\\
$^3$Department of Electrical Engineering and Computer Science,
University of Michigan, 2260 Hayward Street, Ann Arbor, MI
48109-2121, USA}
\date{\today}

\begin{abstract}
Understanding the nature of multipartite entanglement is a central mission of quantum information theory.
To this end, we investigate the question of tripartite entanglement convertibility.  We find that there exists no easy criterion to determine whether a general tripartite transformation can be performed with a nonzero success probability and in fact, the problem is NP-hard.  Our results are based on the connections between multipartite entanglement and tensor rank (also called
Schmidt rank), a key concept in algebraic complexity theory.  Not only does this relationship allow us to characterize the general difficulty in determining possible entanglement transformations, but it also enables us to observe the previously overlooked fact that {\em the Schmidt rank is not an additive entanglement measure}.  As a result, we improve some best known transformation rates between specific tripartite entangled states.  In addition, we find obtaining the most efficient algorithm for matrix multiplication to be precisely equivalent to determining the optimal rate of conversion between the Greenberger-Horne-Zeilinger state and a triangular distribution of three Einstein-Podolsky-Rosen states.

\end{abstract}


\maketitle

One of the greatest discoveries in quantum physics \cite{EPR} is that a multipartite quantum system
can be in a so-called entangled state. There are an uncountable number of entangled states realizable by any quantum system and a natural question is how they are related to each other --- specifically, if two given states can be converted to each other through {\em local operations and classical communications (LOCC)}, i.e. a protocol in which no quantum information is exchanged among the subsystems. 
Ideally, one would like to have an efficient way to decide whether one entangled state can be transformed into another via LOCC as this question has utmost physical relevance. By nature, entangled states are fragile and highly susceptible to {\em decoherence}, a process in which quantum powers are lost.  LOCC protocols describe the cheapest and experimentally easiest ways to convert entanglement while minimizing decoherence among parties separated by arbitrary distances.

As the notion of probability is inherent to quantum mechanics, the more natural question is with what probability $p$ can $|\phi\rangle$ be converted into $|\psi\rangle$ under LOCC?  For $p=1$, the LOCC transformation is called {\em deterministic} and for a general nonzero $p$, the protocol is called {\em stochastic (SLOCC)}.  Transformations of the latter form are written as $|\phi\rangle\slocc|\psi\rangle$.  For bipartite systems, the problem is completely solved.  Nielsen has provided necessary and sufficient conditions for whether two states are deterministically convertible \cite{nielsen-99-83}.  Probabilistically, any bipartite state $|\phi\rangle$ can be transformed into $|\psi\rangle$ if and only if the matrix rank of the reduced density operator of $|\phi\rangle$ is greater 
than that of $|\psi\rangle$.  Furthermore, Vidal \cite{vidal-99-83} has derived a simple formula that gives the optimal probability for conversion.  

When the number of subsystems is greater than two, the situation becomes much more complicated.  No longer can SLOCC convertibility be determined by examining the ranks of the reduced density matrices of the initial and final states.  For example, a system of three qubits can be partitioned into six equivalence classes defined by SLOCC convertibility between states in the same class \cite{DVC-2000}. However, two of the classes are indistinguishable by examining the ranks of each subsystem's density matrix.  The states $|GHZ\rangle=\frac{1}{\sqrt{2}}(|000\rangle+|111\rangle)$ and $|W\rangle=\frac{1}{\sqrt{3}}(|001\rangle+|010\rangle+|100\rangle)$ are representatives of each class respectively.  Progress toward understanding SLOCC convertibility among four qubit states has also been made as the state space for these systems can be partitioned into nine different families of equivalence classes \cite{verstraete-2002-65}.  Nevertheless, for both three and four qubit systems, the separation into SLOCC equivalence classes does not provide a solution for determining whether two states are SLOCC related simply because one must first determine to which classes the states belong.  

In this letter, we ask whether there is some relatively simple criterion for determining the convertibility of arbitrary tripartite states like there is for bipartite states.  As a complete solution to the convertibility problem should be able to determine whether one state can be transformed into another with a nonzero probability, we focus our attention on the class of SLOCC protocols to judge the difficulty of the complete problem.  Ultimately we find that no simple criterion exists for testing the possibility of a general tripartite entanglement transformation.  In addition, through the course of investigating this problem many other interesting results are obtained concerning specific tripartite transformation rates.  The novel conversion rates are derived in part from our observation that the Schmidt measure (to be defined below) is {\em not an additive quantity}, something previously thought to be true \cite{eisert-2001-64}.  We now summarize our main findings:

Denote by $\TEPR$ the unnormalized tripartite state where any two parties share an (unnormalized) EPR state $|\Phi\rangle=|00\rangle+|11\rangle$:
\begin{align}
|\Phi^3\rangle&=|\Phi\rangle_{AB}|\Phi\rangle_{AC}|\Phi\rangle_{BC}\notag\\&=\Big(|00\rangle_A|00\rangle_B+|10\rangle_A|10\rangle_B\Big)|00\rangle_C\notag\\&+\Big(|00\rangle_A|01\rangle_B+|10\rangle_A|11\rangle_B\Big)|01\rangle_C\notag\\&+\Big(|01\rangle_A|00\rangle_B+|11\rangle_A|10\rangle_B\Big)|10\rangle_C\notag\\&+\Big(|01\rangle_A|01\rangle_B+|11\rangle_A|11\rangle_B\Big)|11\rangle_C.
\end{align} 

\begin{theorem}\label{thm:main}
\mbox{}
\begin{itemize}
\item[(a)] For any general tripartite $|\phi\rangle$ and $|\psi\rangle$, determining whether $|\phi\rangle$ can be obtained from $|\psi\rangle$ by SLOCC is NP-hard.
\item[(b)] $\GHZ^{\otimes 3}\slocc\W^{\otimes 2}$.
\item[(c)] $\GHZ^{\otimes 17}\slocc|\Phi^3\rangle^{\otimes 6}$.
\item[(d)] Let $\lambda=inf\{u:|GHZ\rangle^{\otimes \lfloor un\rfloor}\slocc|\Phi^3\rangle^{\otimes n}$ {\em for sufficiently large} $n\}$.  Then $\lambda$ is 
precisely the {\em exponent for matrix multiplication}, i.e., 
the smallest real number $\omega$ such that two $N$ by $N$ matrices can be multiplied
with $O(N^{\omega})$ number of multiplications between linear functions on entries of the first matrix
and linear functions on entries of the second matrix.
\end{itemize}
\end{theorem}
 
Previously, only one copy of the W state is known to be convertible from three copies of GHZ and result (b) provides an improvement to this rate.  Transformation (c) is important because it reveals that the three-party EPR extraction rate from GHZ is greater than one, a previously unknown possibility.  Result (d) shows that existing lower and upper bounds for matrix multiplication translate to lower and upper bounds on the optimal conversion rate between groups of the given states.  This connection implies that finding the GHZ to three-party EPR conversion rate is highly difficult since the complexity of multiplying matrices is one of the most challenging open problems in computation theory.  

Our main technical tool is tensor rank, a key concept in algebraic complexity theory \cite{Burg97}
that has also been used to measure multipartite entanglement under the synonymous names of Schmidt rank and Schmidt measure \cite{eisert-2001-64}.
The {\em tensor rank}
of a multipartite state  $|\phi\rangle\in H_1\otimes H_2\otimes \cdots \otimes H_n$, denoted by $rk(|\phi\rangle)$,
is the minimum
number $r$ such that there exists $|\phi_j\rangle_i\in H_i$, $1\le j\le r$ and
\[ |\phi\rangle=\sum_{j=1}^r \bigotimes_{i=1}^n |\phi_j\rangle_i.\]
The quantity $log_2(rk(|\phi\rangle)$ is called the {\em Schmidt measure} of $|\phi\rangle$,
denoted by $sch(|\phi\rangle)$.

Tensor rank has been used in algebraic complexity theory as it captures the complexity of computing a set of bilinear maps \cite{Burg97} and in particular the multiplicative complexity of multiplying two matrices.  A set of bilinear maps are polynomials with respect to two distinct groups of indeterminates.  The {\em multiplicative complexity} of the set is the minimum number of multiplications between the two groups required to evaluate all the polynomials.  The multiplication of two $N\times N$ matrices produces a set of $N^2$ bilinear maps, one for each entry in the $N\times N$ product.  The complexity of $N\times N$ matrix multiplication is denoted by $\mu(N,N)$ and the current best upper and lower bounds for $\mu(N,N)$ are $O(N^{2.36})$ and $\frac{5}{2}N^2-3N$ respectively \cite{CS&W-87, Blaser-2005}.  The complexity of matrix multiplication is also expressed as $\mu(N,N)=O(N^\omega)$ where $\omega$ is called the {\em exponent for matrix multiplication} and defined as the smallest real number such that an algorithm exists for multiplying two $N\times N$ matrices using $O(N^\omega)$ multiplications.  While $\omega$ is hypothesized to be two, determining the validity of this conjecture is a major open problem in computational science.  For more details, a good reference is chapter 28 of \cite{CLR-algorithms}.  

Tensor rank analysis has already shown to be valuable in 
quantum information as it is the distinguishing property between the $|GHZ\rangle$ and $|W\rangle$ equivalence classes of three qubits \cite{DVC-2000, brylinski-2000}.  It has also been useful in characterizing the entanglement in graph states \cite{hein-2004-69} as well as studying the distinguishability of states by separable operations \cite{duan-2007}.  
An important property of the tensor rank is that it cannot increase under SLOCC:
\begin{proposition}\upshape{\cite{lo-1997}}\label{prop:monotone}
\textit{If $|\phi\rangle\slocc|\psi\rangle$ then $\rk(|\phi\rangle)\ge\rk(|\psi\rangle)$}.
\end{proposition}

Through Proposition~\ref{prop:monotone}, the monotonic nature of the tensor 
rank makes studying it physically worthwhile. 
Unfortunately, determining the rank of an arbitrary state is a very difficult problem \cite{Haastad-1990} which is ultimately why there is no simple convertibility test applicable to all tripartite transformations. 
However, in some special cases it is possible to calculate the tensor rank or at least determine some useful bounds.
In this Letter, we establish our main results described above by examining 
the ranks of certain tripartite states. We prove the following where each statement is in
one-to-one correspondence with the main results stated earlier.

\begin{lemma}
\label{lm:main}
\mbox{}
\begin{itemize}
\item[(a')] $|\phi\rangle\in H_A\otimes H_B\otimes H_C$ can be SLOCC converted from state $\frac{1}{\sqrt{n}}\sum_{i=1}^N|i\rangle_A|i\rangle_B|i\rangle_C$ if and only if $rk(|\phi\rangle)\leq N$.
\item[(b')] $\rk(\WW)\le 8$. 
\item[(c')] $\rk(|\Phi^3\rangle)=7$.
\item[(d')] $rk(\TEPR^{\otimes n})$ is the multiplicative complexity for multiplying two $2^n\times 2^n$ matrices.
\end{itemize}
\end{lemma}
These results immediately indicate that when many copies of a state are considered, 
the tensor rank does not necessarily scale proportionately. As a result, impossible SLOCC 
transformations between individual sates may be possible when bulk quantities are 
considered. 

Extending Lemma 1 to prove Theorem 1 is straightforward.  It follows from item (a') that, given a tripartite tensor $|\phi\rangle$ and a number $k$, deciding if $\rk(|\phi\rangle)\le k$ can be reduced
to the question of whether $\sum_{i=1}^k|i\rangle_A|i\rangle_B|i\rangle_C\slocc |\phi\rangle$.
The former problem is shown to be NP-hard by H{\aa}stad \cite{Haastad-1990}, thus the latter is also 
NP-hard (item (a)).

Results (b), (c), and (d) follow directly from applying (a') to (b'), (c'), and (d') respectively.  The 17 to 6 conversion ratio of (c') is important because 6 copies of $|\Phi^3\rangle$ is a total of 18 EPR pairs.  Thus, the stochastic EPR distillation rate from multiple copies of $|GHZ\rangle$ is greater than 1.  In fact (d) shows that this rate can be further improved as the upper bound for $\omega$ is lowered.  However, the distillation is specific in that the EPR pairs must be shared among all three parties.  Indeed, if the EPR pairs are held by just two parties, $rk(|\Phi\rangle^{\otimes n})=2^n$ so the EPR distillation rate from $n$ copies of $|GHZ\rangle$ equals 1.  The related problem of EPR distillation from the W state has recently been studied in \cite{Lo-2007}.  There, the authors show that for a single W state, the probability of extracting an EPR state via LOCC is not only higher if one does not specify which two parties share the state, but it can also be made arbitrarily close to one.

From (d) and the lower bound on $\mu(2^n,2^n)$, it follows that $2n$ copies of GHZ cannot be converted into $n$ copies of $|\Phi^3\rangle$ with a nonzero probability.  This result is stronger than the one derived in \cite{linden-1999} where the authors demonstrate the impossibility of $|GHZ\rangle^{\otimes 2n}\rightarrow|\Phi^3\rangle^{\otimes n}$ under deterministic LOCC.  What is most interesting is that the authors prove the impossibility strictly through entropy arguments.  Here, we obtain the same conclusion using tools of algebraic complexity theory. On the surface these two lines of attack appear to be unrelated, but the similarity in both results suggests that the two may be deeply connected.

Now we turn to prove Lemma 1.  We will work with unnormalized states below since any overall factor does not affect the tensor rank.  For any $|\phi\rangle\in H_A\otimes H_B\otimes H_C$, let $\rho_{AB}$ denote Alice and Bob's subsystem obtained by taking the partial trace $Tr_C(|\phi\rangle\langle\phi|)$.  As $\rho_{AB}$ is a positive operator, it has a spectral decomposition $\rho_{AB}=\sum_{k=1}^mp_k|\psi_k\rangle\langle\psi_k|$ where $0<p_k\le 1$.  The vector span of $\{|\psi_k\rangle:1\le k\le m\}$ is called the support of $\rho_{AB}$ and denoted by $supp(\rho_{AB})$.  To proceed, we need the following simple equivalent characterization of a tripartite state's tensor rank.   

\begin{lemma}\label{lm2}  Suppose $|\phi\rangle\in H_A\otimes H_B\otimes H_C$.  The tensor rank of $|\phi\rangle$ equals the minimum number of product states in $H_A\otimes H_B$ whose linear span contains the support of $\rho_{AB}=Tr_C(|\phi\rangle\langle\phi|)$.
\end{lemma}
\begin{proof}[\bf Proof.]
Let $k$ denote $rk(|\phi\rangle)$.  Suppose that the span of $r$ product states $\{|\alpha_j\rangle|\beta_j\rangle:1\le j\le r\}$ contain $supp(\rho_{AB})$.  Let $|\phi\rangle=\sum_{i=1}^m|i\rangle_{AB}|i\rangle_C$ be a Schmidt decomposition of $|\phi\rangle$.  Each $|i\rangle_{AB}$ belongs to $supp(\rho_{AB})$ and thus $|i\rangle_{AB}=\sum_{j=1}^r\lambda_{i,j}|\alpha_j\rangle|\beta_j\rangle$.  Regrouping the $|i\rangle_C$ according to the $r$ product states gives $r\ge k$.  On the other hand, consider a ``minimal'' decomposition $|\phi\rangle=\sum_{i=1}^k|a_i\rangle|b_i\rangle|c_i\rangle$.  Then $\rho_{AB}=\sum_{i,j=1}^k|a_i\rangle|b_i\rangle\langle c_j|c_i\rangle\langle a_j|\langle b_j|$ and hence $k\ge r$.
\end{proof}

Using Lemma~\ref{lm2}, the general procedure for determining tensor rank is now straightforward.  Write $|\phi\rangle=\sum_{i=1}^m|i\rangle_{AB}|i\rangle_C$ where the $\{|i\rangle_C:1\le 1\le m\}$ are orthonormal and then determine the minimum number of product states needed to contain the $\{|i\rangle_{AB}:1\le i\le m\}$.  This question can be rephrased in another way by mapping each $|i\rangle_{AB}$ to a bilinear form $f_i$ from the ring of indeterminates $C[\{a_j\},\{b_j\}]$ where each $a_j$ $(b_j)$ is in a one-to-one correspondence with a basis vector from $H_a$ $(H_b)$.  Product states in $H_a\otimes H_b$ correspond to a product of linear forms from $C[\{a_j\}]\times C[\{b_j\}]$ which we refer to as a {\em non-scalar} multiplication.  Thus, we obtain the following fact:
\begin{fact}
The minimum number of product states that contain the $\{|i\rangle_{AB}:1\le i\le m\}$, and hence the tensor rank of $|\phi\rangle$, is the same number of non-scalar multiplications $M_k=(\sum_{j=1}^{n_a}\alpha_{k,j}a_j)\times(\sum_{j=1}^{n_b}\beta_{k,j}b_j)$ needed to calculate the $\{f_i:1\le i\le m\}$.  
\end{fact}
We now use the technique outlined above to study the tensor rank of certain tripartite states.

\begin{proof}[\bf Proof of Lemma 1.]
(a'):  For $\sum_{i=1}^N|i\rangle_A|i\rangle_B|i\rangle_C$, the support of $\rho_{AB}$ is spanned by $N$ product states.  Thus by Prop. 1 and Lemma 2, a necessary condition for the given transformation is $rk(|\phi\rangle)\leq N$.  Now suppose that $|\phi\rangle=\sum_{i=1}^k|a_i\rangle|b_i\rangle|c_i\rangle$ where $k\leq N$.  Since $\{|i\rangle_A:1\le i\le N\}$ is an orthonormal set, we can define 
the linear operator $A$ by $A|i\rangle_A=\begin{cases}|a_i\rangle, &1\leq i\leq k\\0,&k<i\leq N\end{cases}$.  Similarly, operators $B$ and $C$ can be constructed.  As noted in \cite{DVC-2000}, the existence of such operators is sufficient for an SLOCC protocol since $|\phi\rangle$ will be obtained when Alice performs the local measurement $\{\frac{A}{||A||},\sqrt{I_A-\frac{1}{||A||^2}A^\dagger A}\}$ and similarly for Bob and Charlie.  Note that (unnormalized) $|GHZ\rangle^{\otimes n}$ can be expressed as $\sum_{i=1}^{2^n}|i\rangle_A|i\rangle_B|i\rangle_C$.  (b'): One can verify by direct computation that $|W\rangle^{\otimes 2}$ expands as:
\begin{align}
\label{W2}
\big(|11\rangle_A|00\rangle_B+|10\rangle_A|01\rangle_B+|01\rangle_A|10\rangle_B+|00\rangle_A|11\rangle_B\big)&|00\rangle_C\notag\\
+\big(|10\rangle_A|00\rangle_B+|00\rangle_A|10\rangle_B\big)&|01\rangle_C\notag\\
+\big(|01\rangle_A|00\rangle_B+|00\rangle_A|01\rangle_B\big)&|10\rangle_C\notag\\
+\big(|00\rangle_A|00\rangle_B\big)&|11\rangle_C.
\end{align}  The structure of $|W\rangle^{\otimes 2}$ becomes more manageable when working with its corresponding bilinears $f_i$ since they can be succinctly expressed through the matrix multiplication
$$\begin{pmatrix}
f_{00}\\
f_{01}\\
f_{10}\\
f_{11}\\
\end{pmatrix}
=
\begin{pmatrix}
a_{11}&a_{10}&a_{01}&a_{00}\\
a_{10}&0&a_{00}&0\\
a_{01}&a_{00}&0&0\\
a_{00}&0&0&0\\
\end{pmatrix}
\cdot
\begin{pmatrix}
b_{00}\\
b_{01}\\
b_{10}\\
b_{11}\\
\end{pmatrix}.
$$
We make use of the following identity \cite{Fiduccia-1972-CCC}, where a ``$\cdot$''
means a $0$ entry:
\begin{widetext}
\begin{align}
\label{w2decomp}
\begin{pmatrix}
a_{11}&a_{10}&a_{01}&a_{00}\\
a_{10}&0&a_{00}&0\\
a_{01}&a_{00}&0&0\\
a_{00}&0&0&0\\
\end{pmatrix}
&=
\begin{pmatrix}
a_{10}&a_{10}&\cdot&\cdot\\
a_{10}&a_{10}&\cdot&\cdot\\
\cdot&\cdot&\cdot&\cdot\\
\cdot&\cdot&\cdot&\cdot\\
\end{pmatrix}
+
\begin{pmatrix}
a_{01}&\cdot&a_{01}&\cdot\\
\cdot&\cdot&\cdot&\cdot\\
a_{01}&\cdot&a_{01}&\cdot\\
\cdot&\cdot&\cdot&\cdot\\
\end{pmatrix}
+
\begin{pmatrix}
a_{00}&\cdot&\cdot&a_{00}\\
\cdot&\cdot&\cdot&\cdot\\
\cdot&\cdot&\cdot&\cdot\\
a_{00}&\cdot&\cdot&a_{00}\\
\end{pmatrix}\notag\\
&+
\begin{pmatrix}
\cdot&\cdot&\cdot&\cdot\\
\cdot&a_{00}&a_{00}&\cdot\\
\cdot&a_{00}&a_{00}&\cdot\\
\cdot&\cdot&\cdot&\cdot\\
\end{pmatrix}
+
\begin{pmatrix}
a_{11}-a_{10}-a_{01}-a_{00}&\cdot&\cdot&\cdot\\
\cdot&-a_{10}-a_{00}&\cdot&\cdot\\
\cdot&\cdot&-a_{01}-a_{00}&\cdot\\
\cdot&\cdot&\cdot&-a_{00}
\end{pmatrix}.
\end{align}
\end{widetext}
Note that rank one matrices require only one non-scalar multiplication: $\begin{pmatrix} a_i&a_i\\a_i&a_i\end{pmatrix}\cdot\begin{pmatrix} b_1\\b_2\end{pmatrix}=\begin{pmatrix} a_i(b_1+b_2)\\a_i(b_1+b_2)\end{pmatrix}$, while any $n\times n$ diagonal matrix requires $n$ multiplications: $\begin{pmatrix} \lambda_1&{}&{}&{}\\{}&\lambda_2&{}&{}\\{}&{}&\lambda_3&{}\\{}&{}&{}&\lambda_4\end{pmatrix}\cdot\begin{pmatrix}b_1\\b_2\\b_3\\b_4\end{pmatrix} =\begin{pmatrix}\lambda_1b_1\\\lambda_2b_2\\\lambda_3b_3\\\lambda_4b_4\end{pmatrix}$.  Hence, a total of eight non-scalar multiplications is sufficient to compute each $f_i$.  These multiplications correspond to product states that contain $supp(Tr_C(|W\rangle\langle W|^{\otimes 2}))$.  By Lemma 2 then, $rk(|W\rangle^{\otimes 2})\leq 8$.  In fact, expansion \eqref{w2decomp} gives the eight product states that contain $supp(Tr_C(|W\rangle\langle W|^{\otimes 2}))$ enabling us to rewrite Alice and Bob's vector attached to $|i\rangle_C:i\in\{00,01,10,11\}$ in \eqref{W2} as a combination of these eight states.  To our knowledge, this is the first observed non-additivity of the Schmidt measure for pure states.
(c'):  Up to a local unitary transformation on Alice's part, the corresponding bilinear forms of $|\Phi^3\rangle$ match the set of polynomials obtained when multiplying two $2\times 2$ matrices:
\begin{equation}
\label{mm}
\begin{pmatrix}
a_{00}&a_{01}\\
a_{10}&a_{11}\\
\end{pmatrix}
\begin{pmatrix}
b_{00}&b_{01}\\
b_{10}&b_{11}\\
\end{pmatrix}
=\begin{pmatrix}
f_{00}&f_{01}\\
f_{10}&f_{11}\\
\end{pmatrix}.
\end{equation} 
An algorithm for obtaining the $f_i$ using only seven multiplications was discovered by Strassen \cite{Strassen69} and later proven to be optimal by Winograd \cite{winograd-1971-4}.  These seven non-scalar multiplications correspond to a minimum number of product states containing $supp(Tr_C(|\Phi^3\rangle\langle\Phi^3|))$ and so $rk(|\Phi^3\rangle)=7$.  As an eight term expansion for $|W\rangle^{\otimes 2}$ can easily be obtained from expansion \eqref{w2decomp}, it is straightforward to find a seven term expansion of $|\Phi^3\rangle$ from Strassen's algorithm given in \cite{Strassen69}.  Since the explicit expressions are not of primary interest here, we omit the calculations.  (d'): By taking multiple tensor products of the matrices in \eqref{mm}, we see that for $n$ copies of $|\Phi^3\rangle$, the corresponding polynomials are represented by $2^n\times 2^n$ matrix multiplication.  Hence, $rk(\TEPR^{\otimes n})$ is the complexity of this operation.
\end{proof}

In conclusion, we have found that no easy test exists for determining whether two general tripartite states are probabilistically convertible.  The difficulty arises because any general solution involves a tripartite tensor rank computation.  As a result, one must consider tripartite transformations on a case-by-case basis.  In this letter we have done this for special states in which the tensor rank has already been studied or can be calculated with mild effort.  Performing this analysis led to an improved GHZ state to W state SLOCC transformation rate as well as the first demonstration of obtaining EPR pairs from GHZ states at a rate greater than one with a nonzero probability.

The connection between tensor rank and entanglement transformation is perhaps most beautifully exemplified by the equivalence of matrix multiplication complexity and the optimization of stochastic EPR distillation from many copies of GHZ.  This relationship opens many avenues of further research as the techniques of algebraic complexity theory might teach us more about the nature and limitations of SLOCC transformations.  Conversely, constructing explicit SLOCC entanglement transformations using results in quantum information may be useful to obtain bounds for the multiplicative complexity of a particular set of bilinear forms.

This work was partially supported by the National
Science Foundation of the United States under Awards~0347078 and
0622033. R. Duan was partially supported by the National Natural Science Foundation of China (Grant Nos. 60702080, 60736011, and 60621062), the FANEDD under Grant No. 200755, and the Hi-Tech Research and Development Program of China (863 project) (Grant No. 2006AA01Z102).

\bibliography{QuantumBib}

\end{document}